# INVESTIGATING CONFORMATION CHANGES AND NETWORK FORMATION OF MUCIN IN JOINTS FUNCTIONING IN HUMAN LOCOMOTION

**Natalia Kruszewska and Piotr Bełdowski**

UTP University of Science and Technology, Institute of Mathematics and Physics,

Kaliskiego 7, PL-85796 Bydgoszcz, Poland

**Krzysztof Domino**

Institute of Theoretical and Applied Informatics, Polish Academy of Sciences,

Bałtycka 3 , PL-44100 Gliwice, Poland

**Kanika D Lambert**

Delaware State University, Department of Biological Sciences,

1200 N Dupont Hwy, Dover, DE, 19901, USA

**ABSTRACT**

Many different processes take place to facilitate lubrication of the joints functioning in human locomotion system. The main purpose of this is to avoid destroying the articular cartilage. Viscoelastic properties of the joints system are very sensitive on both temperature and concentration changes because of the change in conformation presented in the system proteins and protein network formation. We are searching for an answer to the question on how changes in temperature and concentration influence the conformational entropy of mucin protein which is a part of one of the key components, lubricin, which is believed to be responsible for gel formation inside synovial fluid. We are using molecular dynamic technique to obtain the information about dihedral ($\phi$, $\psi$) angles of the mucin during protein self-assembly by means of the computer simulation with a time duration up to 50 ns, parameterized by six temperatures ranged between 300-315 K, and six concentrations 10.68-267.1 g/L. The results show that between c3 and c4 (160 g/L and 214 g/L) a transition exists where crowding begins affecting the dynamics of protein network formation. In such a concentration ranges mucin has a chance to change the frictional properties of the system. Simultaneously there were no significant changes in conformations of the mucin's molecules even after they created networks. The temperature changes also did not affect much of mucin's conformations but it introduced slightly modifications in dihedral angles and after some critical value T=306 K it changed conformational entropy trend from decreasing to raising.

## 1. INTRODUCTION

To provide smooth and stable locomotion, humans and animals are equipped with biomechanically functioning joints, in which one of the most important elements is articular cartilage (AC). AC (composed of chondrocytes and an extracellular matrix containing many various biopolymers, cations and water) is a form of soft tissue that covers surfaces of bones which are separated by a lubricant called synovial fluid (SF). The constituent components of SF are Hyaluronic acid (HA), phospholipids (PL), and lubricin (proteoglycan 4; PRG4) [1]. SF behaves as a dynamic network in which fluctuations of string-shaped components allow entanglements to appear and disappear in the molecular network of the fluid at small time intervals [2]. In the last 50 years there has been ongoing research into the mechanism of lubrication seen within joints [3-5] that avoid a destruction of the joints by lowering a friction coefficient. All the joint components, mentioned earlier, interact with a complicated multiscale, synergistic nature [4,6-9] to create a lubrication phenomenon that is still not well described. It is widely known that a deterioration of AC and SF properties, caused by degenerative joint diseases such as osteoarthritis, results in a painful and difficult joint motion. Thus, it is important to analyze the dynamics of such system. However, due to its complexity, it cannot be studied generally by means of both simulations and theoretical studies. We are only able to examine the role of each component of such complex system, as well as the synergy between these components. In [10] and [11], we focused on the tribological role of HA-PL interactions, the effect also investigated in [12] for its role in tribological surgical adjuvant. To continue building an understandable image of synovial lubrication, we now study the internal dynamics of lubricin within SF. Many different studies point out that lubricin will contribute to the structural organization of synovial fluid. Its role is probably more important in this matter than HA which shows little intermolecular association [13].

It was shown by Chang et al., that solutions of HA-PRG4 play an important role in joint lubrication and wear resistance [14]. It was proven experimentally that lubricin alters the frictional behavior of model hydrophobic and hydrophilic surfaces. They conjectured that lubricin reduces wear by shielding the surfaces from direct contact, in contrast to HA which did not adsorb and did not appreciably alter the friction between their model surfaces. Further, in mentioned experiments, the frictional behavior of a physiologically consistent mixture of lubricin and HA was similar to that of lubricin alone [14]. The significance extends to articular cartilage (AC) systems, whose components possess a dual hydrophobic-hydrophilic nature. The otherwise ordinary affinity of these nonpolar (NP) and polar (P) molecules for similar groups is ultimately what leads to self-ordering seen in AC systems. Lubricin, however, is a heavily O-glycosylated protein macromolecule built from more than 1400 amino acids (AA). It is too complex to be studied by all-atom molecular dynamics (MD), but in a small central region (about 100 AA) mucin-like domains are present [14-16].

Mucin in joint fluid was first isolated by acetic acid precipitation as early as 1846 by Frerichs [17]. Mucins are large glycoproteins which are very widely distributed throughout the different organs of the human body, e.g. in stomach, lungs, respiratory and gastrointestinal tract, liver, kidney, colon, eyes, and ears [18]. They lubricate and protect a large range of epithelial surfaces by forming gel-like mucosae when secreted in large enough concentration [19]. The assumption is that mucin is similarly responsible for the lubricating properties of lubricin. Genetic sequencing has distinguished 22 human mucin genes, denoted as MUC1-MUC22. Although synovial mucin differs from the mucins of epithelial origin [20], it shares many structural similarities with members of the mucins' family which contribute to lubricating properties. In this paper, MUC1 has been chosen as an example of mucin structure as it is most widely distributed in humans. How mucins react on temperature's changes and how they form networks by cross-linking would seem to be of particular importance for its frictional attributes. It can show cartilage's ability to accumulate, transmit, and dissipate mechanical energy during locomotion.

Changes in temperature has a fundamental importance in mechanical behavior of AC. The cartilage stiffness and stress relaxation change with temperature [21]. It can be explained by change in viscoelastic properties of cartilage tissue [22]. The temperature increase in the AC due to partially dissipates the input mechanical energy into heat [23]. Cartilage can be reshaped when heated, for example using laser, RF, or contact heating sources. There are many factors that determines the temperature range for joints. For example, it depends on which joint is under consideration, because joints differ in sizes from fingers, knees, ankles, etc. Another important factor is the season. In the winter, joints react different on the external temperature changes than in the summer [24]. For individuals that have "normal" joints, meaning they do not have any clinical joint problems such as arthritis, the joints temperatures ranging from 31.4˚C to 32.8˚C [24]. These temperatures are lower for individuals with joint problems. In such a case, temperatures of the joints can increase to 36˚C, mainly due to less friction. Thus, the question if temperature changes in such a range described above influence secondary structures and binding abilities of mucin, is of big importance.

The secondary structure and the protein stability of the mucin can be well described by conformational entropy. Changes in the conformational entropy are thought to make substantial contributions to important biochemical processes like protein folding, conformational change, and molecular association (binding) [25]. Scientists found difficulties in experimental measurement of conformational entropy, even though few methods exist to characterize atomic motions, such as NMR relaxation methods [26], AFM-unfolding [27], and neutron spectroscopy which demonstrates the role of conformational entropy in thermal protein unfolding [28]. Theoretical studies are, however, computationally demanding [29-30]. In such a case, computer simulation methods seem to be helpful. There are a few methods of computing the conformational entropy from all-atom MD simulations [25,31-33]. One of them is to calculate it from the distribution of the backbone's dihedral angles ($\phi$, $\psi$) presented in a form of Ramachandran plot. It provides a simplistic view of the conformation of a protein by clustering angles ($\phi$, $\psi$) into district regions inside which specific secondary structures ($\alpha$-Helix, $\beta$-Sheets) can be distinguished.

In this chapter, we are using MD method to study structural (thus, conformational) changes inside mucin protein under various thermodynamic conditions which originate from modifications inside AC and SF components' structures during locomotion. The first objective was to find out if temperature affect the internal structure of mucin. The second, we analyzed if secondary structure of mucin and network formation (binding) depends on concentration of mucin in the SF. Both of the factors can be responsible for changes in mucin's viscoelastic properties which goes in pair with changes in friction inside the joints.

## 2. MATERIAL

Human mucin, MUC1[1], structure (see Fig. 1) was taken from the Protein Data Bank (PDB) [34] and modified using YASARA Structure Software (Vienna, Austria) [35] by adding missing hydrogen atoms. The structure has been obtained from solution by NMR experiment at 303K. Its molecular mass is equal to 12 kDa.

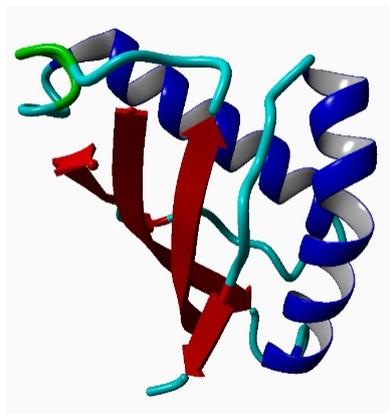

*Figure 1 Ribbon structure of the macromolecule MUC1 drawn using Yasara. It is a complex of two protein chains. The first is built of 66 AA and the second of 55 AA. The structure has been taken from the PDB*

The AMBER03 force field [36] was chosen to evaluate interactions energy between amino acids of mucin due to its proven performance in biophysical systems. Simulations were performed under the following conditions: temperature T=300-315K, pH = 7.0, and 0.9% NaCl aqueous solution (using the four-site model (TIP3P) of water [37], with a time step of 2 fs. Berendsen barostat and thermostat with a relaxation time of 1 fs were used to maintain constant temperature and pressure. The simulation box was given sides of X=130Å, Y=120Å, Z=120Å, and periodic boundary conditions were applied. The mucin molecules were placed in the simulation box in the folded form and after the addition of water molecules the system was minimized for $10^3$ steps with the time step of 2 fs.

---

[1] RCSB PDB code name 2ACM: SEA domain of MUC1

Concentrations of mucin were chosen as follows: $c_0$=10.68 g/L (1 mucin molecule), $c_1$= 53.42 g/L (5 mucin molecules), $c_2$= 106.84 g/L (10 mucin molecules), $c_3$= 160.26 g/L (15 mucin molecules), $c_4$= 213.68 g/L (20 mucin molecules), and $c_5$= 267.1 g/L (25 mucin molecules). Thus, the simulated system probes concentrations in a range encompassing to such that is present in living organisms. The concentrations chosen is such a way makes the model system to accurately simulate mucin both without crowding (too few particles to form a network) and with obvious crowding (stable network formation) in the time range of 0-50 ns.

## 3. METHODS

In order to investigate internal dynamics of the system, all-atom MD simulations of six different temperatures T=300-315K (with ΔT=3K step) in constant concentration $c_0$, and five different concentrations $c_1$-$c_5$ of mucin solution in water in T=310K, were performed. Every case was repeated 10 times to obtain statistically more reliable information. All simulations have been performed in the time range of 0-30 ns in the case of concentration changes and 0-50 ns in the case of temperature changes. We concentrate on computation of two physical parameters: conformational entropy and interactions. All simulations and data analysis have been performed with use of YASARA software and Python programs. The results have been used to describe protein conformation and network formation in (non)crowded systems by means of explaining friction changes in the system.

### 3.1. Ramachandran plot

To get information about conformation of the mucin's chain, a Ramachandran plot has been generated based on knowledge of mucin's atoms positions. The Ramachandran (Rama) plot is a two-dimensional diagram of the dihedral angles (ϕ, ψ) of the protein backbone. It visualizes energetically allowed regions for the dihedral angles ψ against ϕ of amino acid residues in protein chain. The angles (ϕ, ψ) clearly divide into three main distinct regions in the Ramachandran plot where each region corresponds to a specific secondary structure (cf. Fig 2) [38]. We have used the generic type of the Ramachandran plot [39]. It means that the two dihedral angles are measured for the 18 non-glycine, non-proline amino acids. In sequence order, phi (ϕ) is the torsion angle made up by the four atoms: a carbonyl carbon, the connecting α-carbon, an amide nitrogen, and the next carbonyl carbon (C, N, Ca, C). The four atoms which constitute a psi (ψ) are: an amide nitrogen, a carbonyl carbon, an α-carbon, and a second nitrogen (N, Ca, C, N). Such a diagram gives information about what secondary structures are present in the protein and how orderly the structures of all protein are.

### 3.2. Conformational entropy

Backbone conformational entropy has been calculated, based on probability distribution of the (ϕ, ψ), in terms of Shannon entropy which may be expressed as

$$S = -\sum P_i \ln P_i, \qquad (1)$$

where $P_i$ is the fraction of amino acids present in the i-th bin for a specific range of (ϕ, ψ) angles based on histogram taken for Ramachandran plot. In ours case, the Ramachandran's plot was divided into 25 × 25 equally spaced bins. The height and width of each bin is 14.4°. All amino acids in the protein are classified in a specific bin according to the values of their dihedral angles (ϕ, ψ). All probabilities in all bins have been summarized as in Eq. (1).

### 3.3. Hydrogen bond and hydrophobic-polar interactions

In order to check if network is formed between mucins, inter- and intramolecular interactions have been analyzed. Two different type of interactions: hydrogen bonds (HBO) and hydrophobic-polar (HP), have been chosen as network-strength-indicator.

HBO are formed between two oxygen atoms if: (i) the distance between the hydrogen and adjacent oxygen atoms is smaller than 2.6Å; and (ii) the distance between two neighboring oxygen atoms is less than 2.8Å. Hydrogen bond energy, as defined by Eq. (2), is greater than 6.25 kJ/mol (or 1.5 kcal/mol), which is 25% of the optimum value 25 kJ/mol. Thus, only strong bonds (up to 2.6Å) are considered in the analysis. Eq. (2) yields the bond energy in kJ/mol as a function of the Hydrogen-Acceptor distance and two scaling factors [35]

$$E_{HB} = 25 \cdot \frac{2.6 - max(Dist_{H-A}, 2.1)}{0.5} \cdot Scale_{D-A-H} \cdot Scale_{D-A-X}, \quad (2)$$

where the first scaling factor depends on the angle formed by Donor-Hydrogen-Acceptor $Scale_{D-A-H}$, and the second scaling factor is derived from the angle formed by Hydrogen-Acceptor-X $Scale_{D-A-X}$, where the latter X is the atom covalently bound to the acceptor distanced by $Dist_{H-A}$. Both scaling factors vary from 0 to 1 as described in [12].

The HP interaction strength between hydrophobic atoms and hydrogen bond energy are calculated by the algorithm described previously [12].

## 4. RESULTS

Two different MD results sets have been used to bring closer knowledge about mucin molecule properties. The first comes from simulation of single mucin molecule (concentration $c_0$=10.68 g/L) in six temperatures T=300K-315K. The second comes from simulations in T=310K for five concentrations $c_1$- $c_5$(53.42 g/L-267.1 g/L).

### 4.1. Influence of small temperature changes on mucin conformation

Ramachandran plot, originated from MD simulations results of one mucin molecule immersed in water solution with concentration $c_0$=10.68 g/L for six temperatures from 300K to 315K, have been depicted in the Fig. 2. The presented dihedral angles have been computed after 50ns of simulation time. In order to compare the results obtained for mucin, the data from the Top500 structures taken from the scientific report of Lovell group [39] have been taken and put on charts as a background for mucin's angles (ϕ, ψ). The light blue area encloses the "favored" region and 98% of the Lovell's data; the green one encloses the "allowed" 99.95% of the Lovell's data. In the Fig. 2(a), the angles regions for α-helixes (right and left-handed(LF)) and β-sheets have been marked. To make the graph clearer, only two boundary cases of MD results have been shown: for T=300K (red triangles) and T=315K (black dots). In both cases, all mucin's angles presented in Fig. 2 fit well in the three secondary structures' favored or allowed regions. It indicates good quality structure of the mucin – no significant changes in secondary structure can be observed despite changes in temperature. The temperature modification seems to be too small to dramatically change mucin conformation but trend lines show lowering of conformational entropy for temperatures T=300K and T=303K, and increasing of it for the rest of the cases (see, Fig. 3). The changes can be seen in the area of LF α-helixes, and a small number of black dots in 2(b) can be seen outside green α-helixes area. This indicates that a part of α-helixes became less ordered.

In Fig. 3, backbone conformational entropy, computed based on Eq. (1), has been presented. The initial structure of the mucin, taken from PDB to MD simulation, were in native state in T=303K and pH=6.4. That is probably the reason why we can observe two different entropy's behavior below and above 303K. In the lower temperatures (300K-303K), the system found better energetically favorable conformation making angles more fixed in the middle of secondary structure's regions. Angles are less scattered. Increased temperature has provided disorder to the system. Secondary structures start to be less ideal in shapes. Conformational entropy becomes sorted at the end of the simulation time. The *S* values increase accordingly with temperature, even if at the beginning of the simulation the S values had been mixed.

Ramachandran plots, made for probability density of phi (ϕ) and psi (ψ) angles (in histogram form), for six temperatures T=300K-315K, after 50ns of simulation, have been presented in the Fig. 4. Red color intensity is the information about probability density of finding angles (ϕ, ψ) in the specific values ranges described at the axes. The graph shows differences in dihedral angles depend on temperature from more ordered forms (a)-(b) to little more disordered (f). It is of importance that the main secondary structures are still the same as at the beginning of the simulation – no unfolding or refolding process takes place in the studied temperature ranges.

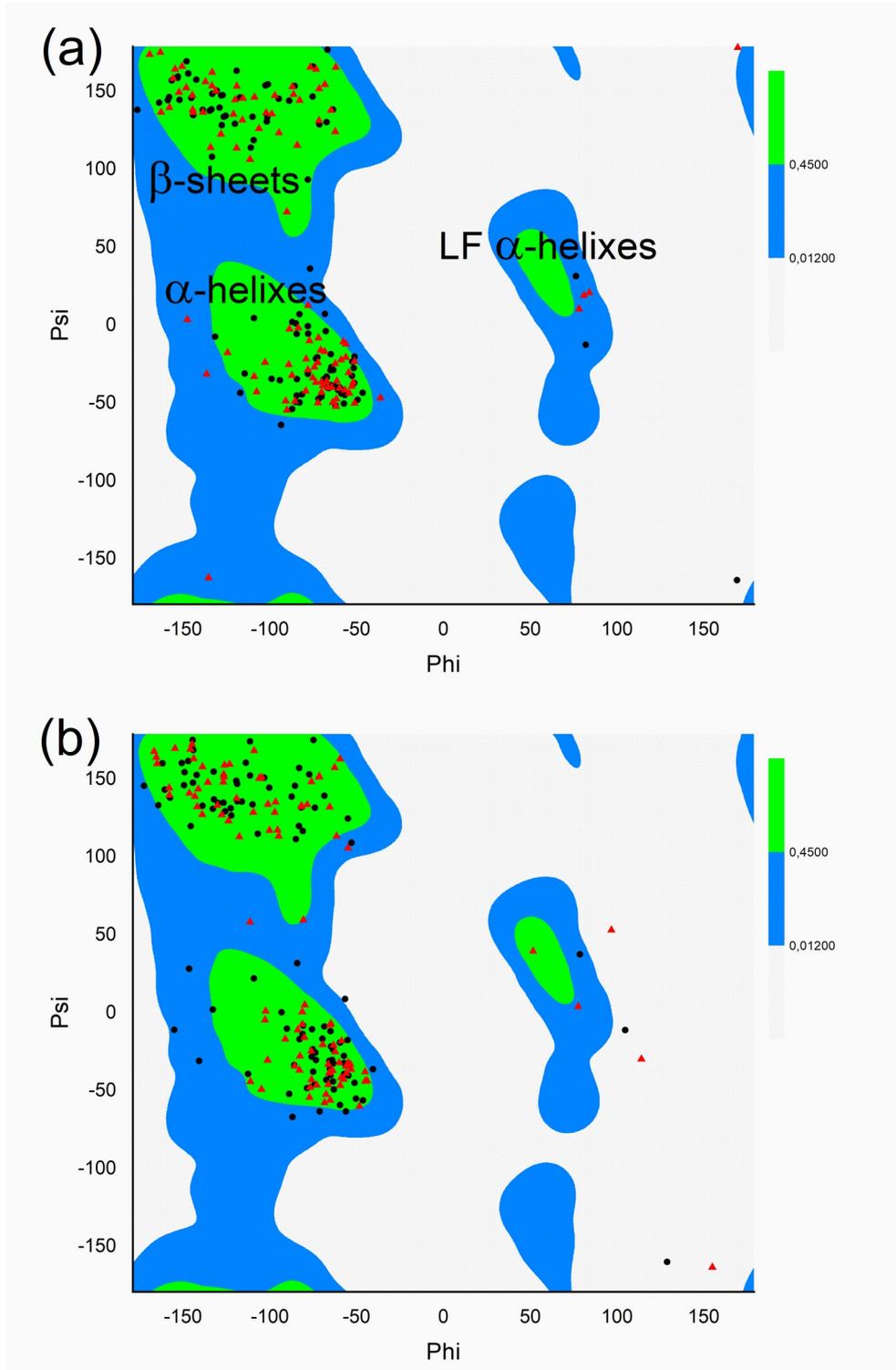

*Figure 2 Ramachandran plots: (a) for T=300K, (b) for T=315K. Red triangles at the beginning of the simulation (0ns), black dots after 50ns of simulation. The blue area encloses the "favored" region and 98% of the Lovell's Top500 data; the green one encloses the "allowed" 99.95% of the Lovell's data [39].*

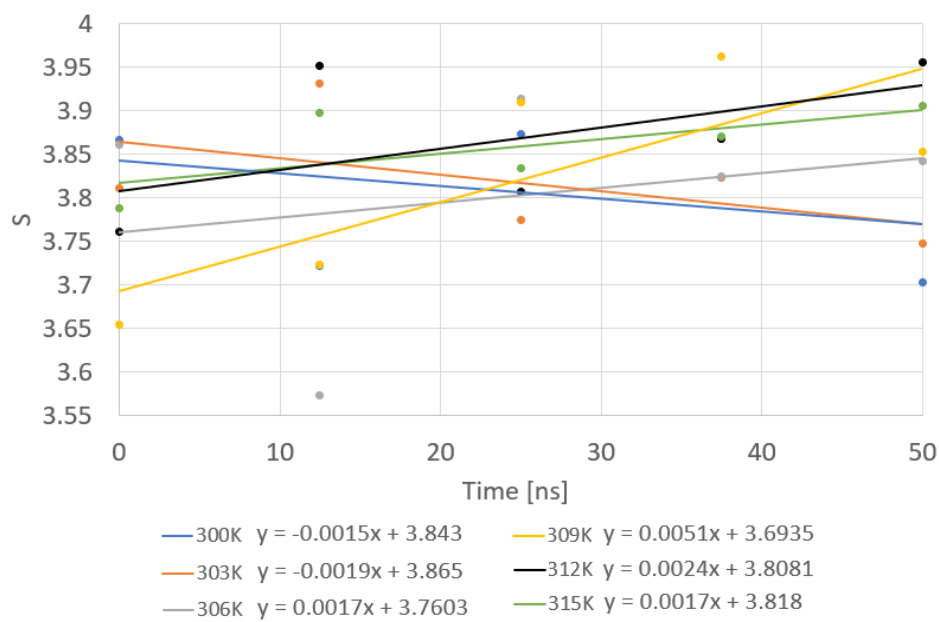

*Figure 3 Trend lines of conformational entropy in a course of time for six different temperatures T=300K-315K*

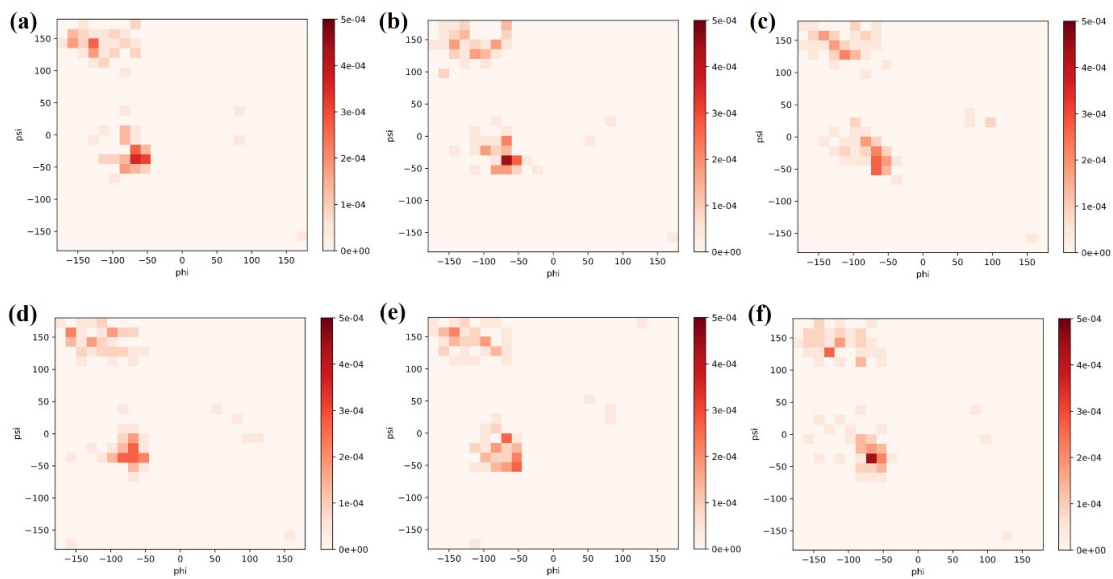

*Figure 4 Ramachandran plot and conformational entropy for single mucin molecule after 50ns of MD simulations in different temperatures: (a) T=300K, S=3.70, (b) T=303K, S=3.74, (c) T=306K, S=3.84, (d) T=309K, S=3.85, (e) T=312K, S=3.96, (f) T=315K, S=3.91*

### 4.2. Influence of concentration changes on mucin conformation

How mucin's concentration influences the conformational entropy, *S*, can be seen in the Fig. 5. The lowering of the conformational entropy as a function of time can be observed in each case but the slope is very small. The values of *S* increasing as the concentration increases but only as a result of having greater number of angles to analyze when greater numbers of mucin molecules are under consideration. The mucins did not significantly change the secondary structures despite of appearance of the intermolecular interactions.

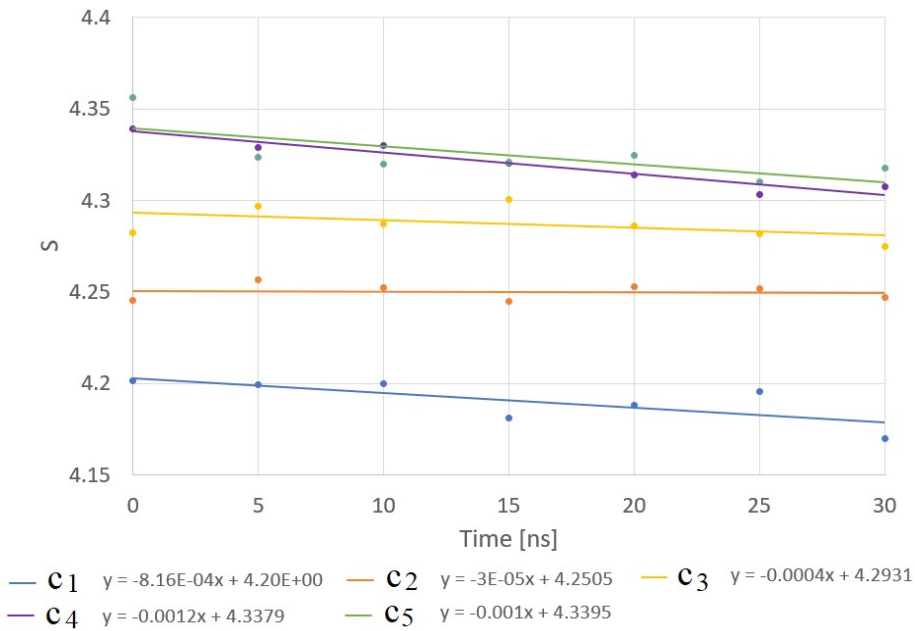

*Figure 5 Conformational entropy in a course of simulation time for five concentrations $c_1$-$c_5$ in T=310K*

Two interactions types: HBO and HP have been analyzed as a function of simulation time for different mucin concentrations. For HP, a number of HP interactions per single mucin molecule has been shown (see, Fig. 6). For HBO we present the total bond energy for a single mucin, which is proportional to the number of bonds (see, Fig. 7). Although the number of bonds alone would suffice, the energy offers additional indication about the energetic regime of the system (cf. Eq(2)).

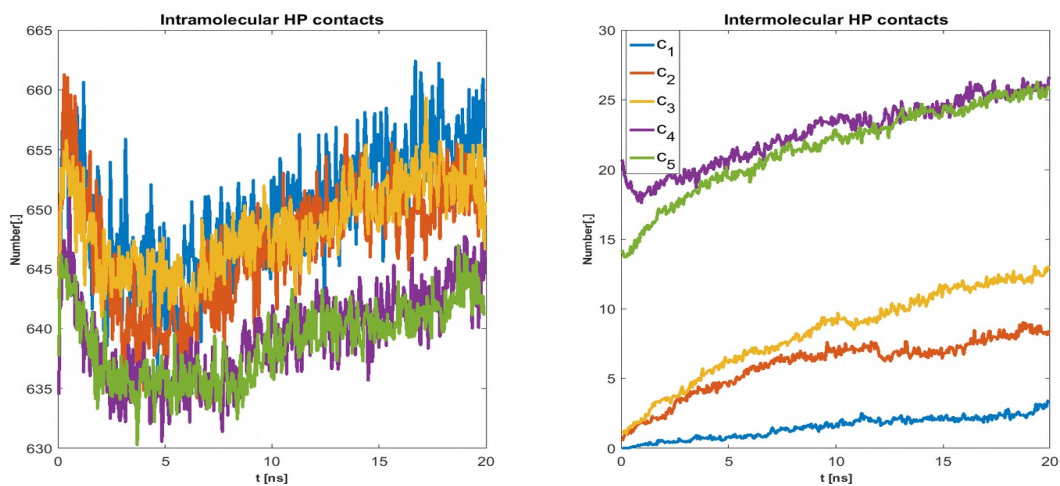

*Figure 6 Number of hydrophobic contacts per single mucin molecule for each concentration. Two cases are presented: (left) Intramolecular; (right) Intermolecular. c1-c5 values are given in Sec. 2*

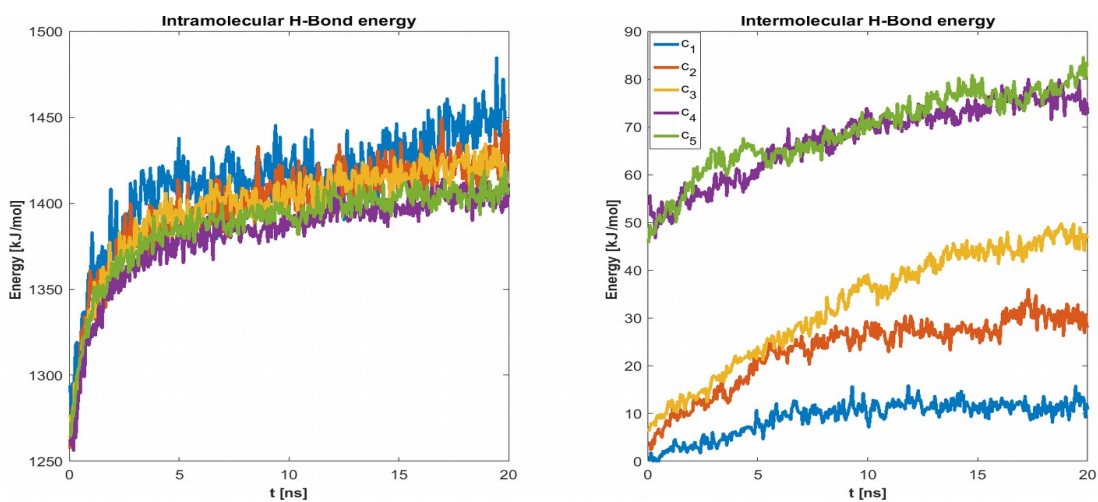

*Figure 7 Total hydrogen bond energy per single mucin molecule for each concentration. Two cases are presented: (left) Intramolecular; (right) Intermolecular. c1-c5 values are given in Sec. 2*

Both results present information about strength of the protein network. Stability of the native proteins is primarily determined by their intramolecular HP interactions, while the stability of protein aggregates depends more on backbone intermolecular HBO interactions [40]. A large difference between the number of inter- and intramolecular interactions for both HP and HBO can be seen. This is due to the fact that the conformations taken from PDB is the protein's folded state, so the number of intramolecular interactions was maximized at the start of simulation. As the proteins aggregated in the course of the simulation the number of intermolecular interactions increased, but more time would be needed for this to reach a maximum. The crowding in higher concentrations likely prevented molecules from finding their lowest energy conformations, while the extra freedom in low concentrations allowed for more fluctuations. That is why for $c_1$ (and also $c_2$) stabilization begins after 10ns for both parameters. Weak networking seems to appear for the first time at concentration $c_3$ as evidenced by the continued increase in intermolecular interactions, but would still be broken with relatively small amounts of energy. This continued increase mirrors those seen for high concentration, just for a smaller absolute number of interactions (a downward shift). The cases $c_4$ and $c_5$ are very similar with almost the same total bond energy and number of HP contacts by the end of the simulation. Thus, increasing the concentration even further would likely not impact the network's mechanical properties.

It is very interesting that in all cases the number of intramolecular HP interactions decreases slightly in the first 5ns of the simulation (about 1.5%). At the same time, increases in the number of intramolecular HP interactions and inter- and intramolecular HBO are seen. This points to a reorganization inside individual single protein chains that is similar across the concentrations. However, this reorganization is not clearly visible on the Ramachandran plot (see, Fig. 8 for $c_1$ concentration – for other concentrations the reorganization similarly is not well visible).

One also notices the relation between parameters inverts when moving from intra- to intermolecular interactions. Increasing concentration favors interactions between molecules while slightly mitigating intramolecular interactions. This, again, is evidence of competition for bonds resulting from increased crowding. In all cases one can notice that the first 10ns of simulation is crucial for network formation - the dynamics of contact creation is initially very fast, but after this time the process slows. In all the interactions charts (cf., Figs 6-7) a conspicuous gap between concentrations is visible. Mainly, between $c_3$ and $c_4$ there is a critical concentration above which a transition seems to occur. More refined concentration values would be necessary to find the optimum conditions for strong network formation in this range - sparse enough to allow conformations to occur but dense enough to provide the necessary cross-linking.

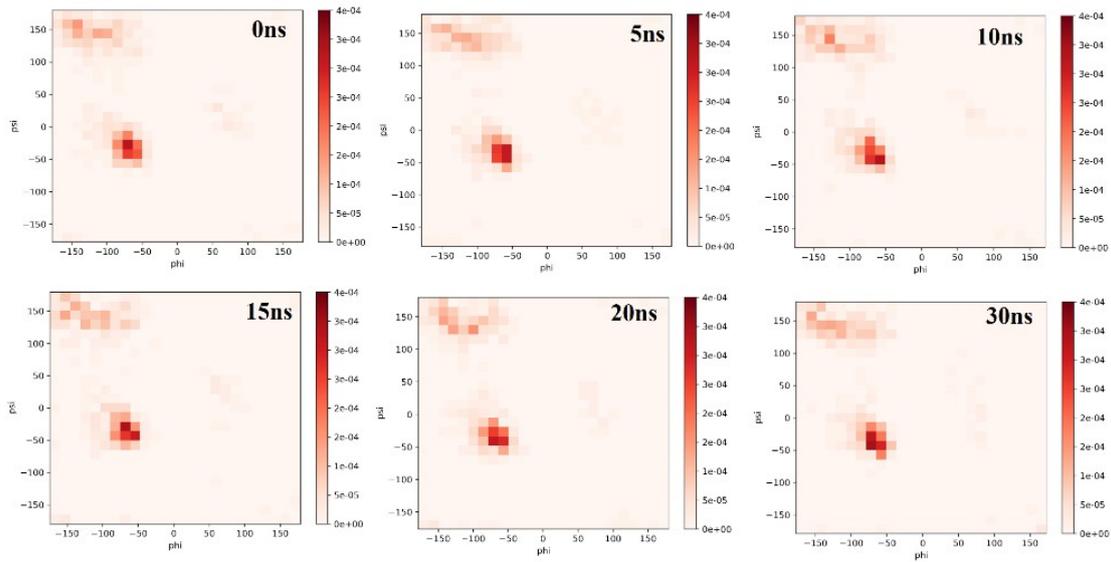

*Figure 8 Ramachandran plot for six time-stamps of MD simulations for $c_1$ concentration*

## 5. CONCLUSIONS

MD simulations of (non)crowded protein system have been performed at different temperatures and concentrations. The results show that one of the most crucial components of the synovial fluid - mucin can only slightly change its conformation depending on both concentration and temperature in the ranges which are present in SF system. Main secondary structures stayed the same but shifts in dihedral angles ($\phi$, $\psi$) as well as changes in the conformational entropy can be clearly seen.

These results can be viewed in terms of facilitated lubrication of articular cartilage system. AC system temperature can increase as a result of rubbing surfaces. Thus, one can say that small differences in a conformational entropy will not disturb a system in a destructive way. Temperature has a big influence on viscosity of fluids but mucin, as the only part of SF seems to not change its conformation in given temperature ranges and in such a dilute environment (with concentration $c_0$). On the other hand, concentration changes can be attributed to regimes of lubrication. From this point of view, lubrication regimes changes from boundary via mixed to hydrodynamic with the decreasing mucin concentration. The concentration is very important for viscoelastic properties of the system which are in turn connected to its lubrication abilities. The results show that between $c_3$ and $c_4$ (160 g/L and 214 g/L) a transition exists where crowding begins affecting the dynamics of protein network formation. The crowding effect is clearly visible through intermolecular HP and HBO interactions but there were no significant changes in conformations of the mucin's molecules even after they created networks (only dihedral angles ($\phi$, $\psi$) were significantly affected). Conformational entropy in most of the cases decreased in time making the system more ordered but in the case of temperatures above 306K it increased introducing little more chaotical structures.

Our next goal is to perform steered molecular dynamics simulations of mucin at different concentrations and external forces applied to understand how the molecule behaves in non-equilibrium conditions.